# Robust Failure Detection Architecture for Large Scale Distributed Systems

Ciprian Dobre*. Florin Pop*, Alexandru Costan*, Mugurel Ionut Andreica*, Valentin Cristea*

*Computer Science Department, Faculty of Automatic Control and Computes Science,
University POLITEHNICA of Bucharest
(e-mails: {ciprian.dobre, florin.pop, alexandru.costan, mugurel.andreica, valentin.cristea}@cs.pub.ro)

**Abstract:** Failure detection is a fundamental building block for ensuring fault tolerance in large scale distributed systems. There are lots of approaches and implementations in failure detectors. Providing flexible failure detection in off-the-shelf distributed systems is difficult. In this paper we present an innovative solution to this problem. Our approach is based on adaptive, decentralized failure detectors, capable of working asynchronous and independent on the application flow. The proposed solution considers an architecture for the failure detectors, based on clustering, the use of a gossip-based algorithm for detection at local level and the use of a hierarchical structure among clusters of detectors along which traffic is channeled. The solution can scale to a large number of nodes, considers the QoS requirements of both applications and resources, and includes fault tolerance and system orchestration mechanisms, added in order to asses the reliability and availability of distributed systems.

## 1. INTRODUCTION

Large scale distributed systems are hardly ever "perfect". Due to their complexity, it is extremely difficult to produce flawless designed distributed systems. While until recently the research in the distributed systems domain has mainly targeted the development of functional infrastructures, today researchers understand that many applications, especially the commercial ones, have some complementary necessities that the „traditional" distributed systems do not satisfy. Together with the extension of the application domains, new requirements have emerged for large scale distributed systems; among these requirements, fault tolerance is needed by more and more modern distributed applications, not only by the critical ones. The clients expect them to work despite faults occurring in such systems.

Although the importance of fault tolerance is today widely recognized and many research projects have been initiated recently in this domain, the existing systems often offer only partial solutions that follow a particular underlying distributed architecture. Traditional fault detection solutions, in particular, fail to work properly in the context of large scale distributed systems because of the large number of involved monitored processes, high probability of messages being lost, the dynamic nature of their topologies and the unpredictable latencies of the message deliveries. The characteristics of large scale distributed systems make fault detection a difficult problem from several points of view. A first aspect is the geographical distribution of resources and users that implies frequent remote operations and data transfers. These lead to a decrease in the system's capability to detect faults, leading to the impossibility to manage correct group communications and consensus. Another problem is the volatility of the resources, which are usually available only for limited periods of time. The system must ensure the correct and complete execution of the applications even in the situations when the resources are introduced and removed dynamically, or when they are damaged. Solving these problems still represents a research problem. The fault detector must be very sensitive to dynamic network conditions. In large scale distributed systems the probability for messages being lost is higher than in case of traditional systems: many resources being contributed means a higher error rate; resources being dynamically introduced or experiencing high loads leads to transitional errors; the background noises can easily be mistaken for faults because of the higher message delivering times. Another problem relates to the flexibility of the applications being executed in such systems. In large scale distributed systems different applications must work concurrently despite of their various objectives, requirements and politics. Such applications inquire different levels of precisions in failure detection. For example, a real-time application requires to be rapidly informed when a process fails. On the other hand, a distributed database could require a higher degree of confidence that a given remote process has failed. A generic failure detector support service must cope with the different QoS requirements coming from various applications.

In this paper we present an innovative solution to solving the requirements involved in obtaining a robust failure detector designed for large scale distributed systems. Our failure detection tool counts for all the aforementioned problems, being particularly designed for highly dynamic large scale distributed systems. Its architecture allows applications to specify different QoS detection levels, while preserving scalability, generality and non-intrusive characteristics.

The rest of this paper is structured as follows. Section 2 presents related work to the problem of designing failure detectors for distributed systems. The next sections present the proposed architecture and key elements of the implementation of a robust failure detector, highlighting the proposed models and protocols. Finally, in Section 5 we present some conclusions and future work.

## 2. RELATED WORK

A failure detector is widely recognized as an oracle that can intelligently suspect processes to have failed (Chandra&Toueg, 1996). In distributed applications, failure detection is generally implemented through the use of directly invoked local services (unreliable local failure detectors).

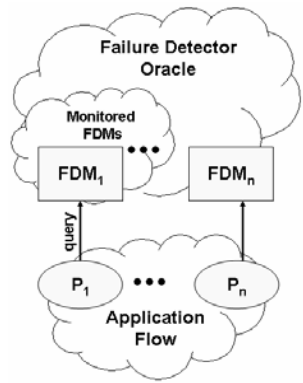

Fig. 1. Federation of unreliable failure detector modules.

The general strategy consists in attaching to each processes of a distributed application a failure detection module (see Figure 1). The failure detection module works asynchronous and independent on the application flow and is responsible with monitoring a subset of the processes in the system and maintaining a list of those it currently suspects to have crashed. A process can query its local failure detector module at any time. Internally, the failure detector module maintains a list of suspect processes that he suspects are crashed. The suspect processes list is permanently updated such that, at any time, new processes can be added and old ones removed. For example, a process suspected to have crashed at time $t$ can be removed from the list at time $t+1$ (it is no longer suspected). The failure detector is considered unreliable (Chandra&Toueg, 1996) because is allowed to make mistakes, to a certain degree. A module can erroneously suspect some correct process (wrong suspicion) or can fail to detect processes that are already crashed. At any given time two failure detector modules may have different lists of processes.

The most common implementation of local failure detection is based on the heartbeat strategy. In this strategy every failure detector module periodically sends a heartbeat message to the other modules, to inform them that it is still alive. When a module fails to receive a heartbeat from another process for a predetermined amount of time (timeout) it concludes the remote process crashed. There is a tradeoff, however, for the timeout values being considered. If the timeout is short then crashes are detected quickly, but there is a high chance of suspecting of being crashed processes that takes a longer time to respond (due to a possible high load for example). Conversely, if the timeout is long, the chance of wrong suspicions if low, but the detection time is deteriorated. This approach does not consider also the heterogeneity of distributed systems. The fact that the timeout is fixed means that the failure detection mechanism is unable to adapt to changing conditions. A long timeout in some systems can turn out to be very short in a different environment.

In the last years there have been many proposals to address some of the problems of ensuring scalable failure detection. In a large scale distributed system, consisting of many nodes, it is impractical to let the failure detection modules monitor each others. An alternative to this consists in arranging processes into an hierarchical structure (such as tree, forest, etc.) along which traffic is channeled. For example, one such solution relies on the use of a two-level hierarchy and is specifically designed for the Globus toolkit (Stelling, *et al*, 1998). However, being a detection scheme based on only a two-level hierarchy, the proposed solution fails to take full advantage of the hierarchical approach and, consequently, do not scale well for large scale distributed systems.

An alternative technique for implementing failure detectors comes in the form of gossip-like protocols. In this approach processes randomly pick partners with whom they exchange their information. The idea is that, with high probability, eventually all processes obtain any piece of information. One of the pioneering works in implementing gossip-style failure detectors is (van Renesse, *et al*, 1998). In their work the authors identified a variant specifically designed for large scale distributed systems: the multilevel gossiping. The idea is to define a multilevel hierarchy using the structure of Internet domains and subdomains as defined by comparing IP addresses. However, the protocol does not work well when a large number of components crash or become partitioned away.

An alternative approach to implementing failure detectors comes in the form of adaptive protocols (Defago, *et al*, 2003). These protocols are designed to adapt dynamically to their environmental and, in particular, adapt their behavior to changing network conditions. A protocol that adjusts the timeout by using the maximum arrival interval of heartbeat messages was proposed in (Fetzer, *et al*, 2001). The protocol assumes a partially synchronous system model, being based on the assumption of a bound on message delays. In (Chen, *et al*, 2002) the authors proposed a different approach based on a probabilistic analysis of network traffic. Their protocol uses arrival times sampled in the recent past to compute an estimation of the arrival of the next heartbeat. The timeout is set according to the estimation and a safety margin, based on application QoS requirements (e.g. upper bound on detection time) and network characteristics (e.g., network load).

A distinctive category of detectors is represented by the accrual failure detectors (Defago, *et al*, 2003)(Defago, *et al*, 2005). The family of accrual failure detectors consists of detector modules that associate, to each of the monitored processes, a real number value that changes over time. One example of an implementation of an accrual failure detector is the φ-failure detector (Defago, *et al*, 2003). The φ-failure detector samples the arrival time of heartbeats and maintains a sliding window of the most recent samples. The window is used to estimate the arrival time of the next heartbeat. A similar approach was also proposed in (Bertier, et al, 2002). However, the proposed failure detectors are poorly adapted to very conservative failure detection because of their vulnerability to message losses. In practice message losses tend to be strongly correlated (i.e., losses tend to occur in bursts). A proposed accrual detector designed to handle this problem is the k-failure detector (Hayashibara, et al, 2004). The k-failure detector takes into account both messages losses and short-lived network partitions, each missed heartbeat contributing to raising the level of defined suspicion according to a predetermined scheme.

An important issue with failure detectors is their scalability. An approach that focuses on the scalability of failure detection was proposed in (Bertier, et al, 2002). However, the proposed system assumes simpler failure semantics such as crash failures. In (Khanna, et al, 2007) the authors proposed a different approach to failure detection, based on stateful identification of the application state.

### 3. SYSTEM MODEL AND DEFINITIONS

*System model*. The system model being considered in this paper is based on the one described in (Defago, et al, 2003). We consider a distributed system consisting of a set of processes $\prod = \{p1,...pn\}$. The system assumes the existence of some global time, known to all processes, the domain of which, denoted by $T$, is an infinitely countable subset of real numbers with no upper bound. We assume that processes always make progress, and that at least $\delta>0$ time units elapse between consecutive steps (the purpose of this being to exclude the case where processes take an infinite number of steps in finite time).

*Failures*. The failure model considered in this paper is based on the model of described in (Chandra&Toueg, 1996). A process can be correct or faulty. A process is faulty if its behaviour deviates from its specification, and a process is correct if it is not faulty. We say that a process fails when its behaviour starts deviating from its specification. Faulty processes never recover.

A failure pattern is a function $F : T \rightarrow 2^{\prod}$, where $F(t)$ is the set of processes that have failed before or at time $t$. The function *correct(F)* denotes the set of correct processes (processes that never belong to failure pattern $F$) while *faulty=$\prod$-correct(F)* denotes the set of faulty processes.

*Failure detectors*. In (Chandra&Toueg, 1996) the authors define failure detectors as a collection of failure detector modules, one attached to each process, that output information on the failure pattern that occurs in an execution. A failure detector module outputs information from a range $R$ of values. A failure detector history $H$ with range $R$ is a function $H: \prod x\ T \rightarrow R$, where $H(p,t)$ is the value output by the failure detector module of process $p$ at time $t$. $H$ is only defined at times when the failure detector module provides an answer to a query; the failure detector module may be queried whenever process $p$ takes a step, and each query eventually results in an answer. The times at which queries $1$, $2$, ... are answered are denoted by the sequence $t_p(1), t_p(2), ...$ Correct processes query their failure detector modules infinitely-many times.

*Binary failure detectors*, such as those defined in (Chandra&Toueg, 1996), output values from the range $R = 2^{\prod}$. If a process is part of the output set, it is suspected to have failed, otherwise it is trusted. An S-transition occurs when a trusted process becomes suspected and a T-transition occurs when a suspected process becomes trusted. The authors in [1] define a class hierarchy of unreliable binary failure detectors, of which of particularly importance is the one called ◊P (eventually perfect). The class is defined by the set of failure detector histories that it permits, as specified by the following two properties of completeness and accuracy.

(STRONG COMPLETENESS) Eventually every faulty process is permanently suspected by all correct processes.

(EVENTUAL STRONG ACCURACY) There is a time after which correct processes are never suspected by any correct process.

*Accrual failure detectors*, such as those defined in (Defago, et al, 2003), output values from a range $R = (R^+)^{\prod}$ (infinite range). In their case history is defined as $H(q,t)(p) = sl_{qp}(t)$. The failure detector module outputs non-negative real values, with each value corresponding to a process and representing the current suspicion level of that process.

*Quality of service metrics for failure detectors*. The authors in (Chandra&Toeug, 1996) define metrics for the quality of service of failure detectors. Quality of service quantifies how fast a failure detector detects failures (completeness) and how well it avoids wrong suspicions (accuracy). For example, assuming the asynchronous model consisting of only two processes, $p$ and $q$, the defined metrics are:

- **Detection time $T_D$** is the time that elapses since the crash of $p$ and until $q$ begins to suspect $p$ permanently.
- **Mistake recurrence time $T_{MR}$** measures the time elapsed between two consecutives mistakes.
- **Mistake duration $T_M$** measures the time it takes for the detector to correct a mistake.
- **Average mistake rate $\lambda_M$** measures the rate at which a failure detector makes mistakes.
- **Query accuracy probability $P_A$** is the probability that the failure detector's output is correct at a random time.
- **Good period duration $T_G$** measures the length of a good period, in which a process is not suspected.

### 4. ARCHITECTURE OF THE FAILURE DETECTOR

The proposed architecture (see Figure 2) is based on the idea of making the fault detector available as a service to applications. such that any distributed application could then use the failure detection capabilities of the failure detector. The architecture is based on the idea of fault detection oracles (Chandra&Toueg, 1996).

The architecture is composed of several failure detection services running inside the distributed system. Each service is responsible with monitoring only a subset of the processes running inside the large scale distributed system and provides, in the form of service functionality, information to upper-level applications regarding the current suspected processes. The architecture is composed of three layers.

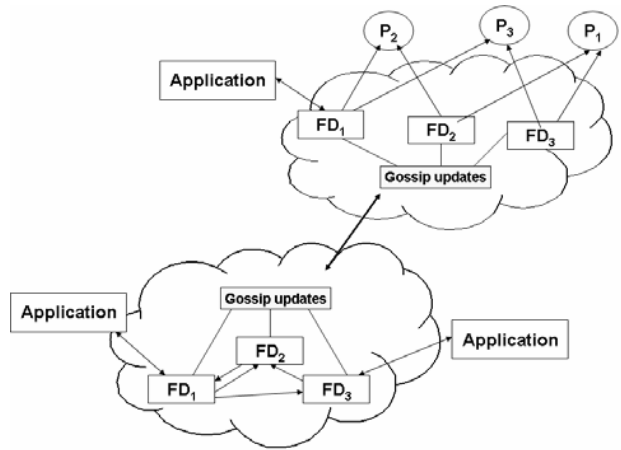

Fig. 2. The architecture of the failure detection service.

A first layer of the architecture is represented by the monitoring function of the detection scheme. At this layer each failure detector is responsible with gathering information about different processes running inside the large scale distributed system. By process we mean either another failure detector or a separate thread running inside a distributed process of a larger application. The monitoring capability is based on sending heartbeat messages to which the remote process must respond. For that, each failure detector sends heartbeat messages to gather information about the state of the distributed processes. More details on the implementation of the monitoring capability of a failure detector process are presented in the next section.

In order to cope with the large nature of the underlying distributed system, the failure detection services are grouped in clusters, each failure detection service being responsible with monitoring all or only a subset of the entire set of processes running inside that particular cluster. The detection scheme uses the advantages of accrual detectors, being able to cope both with the changes in the underlying network, as well as to the dynamic requirements of the applications using the service capabilities of the failure detectors.

However, the failure of a process as detected by a particular daemon can be attributed to several factors: the daemon does not have a direct link anymore with the monitored process, there is an increased background traffic that results in an increase in the time needed for the process to respond back to the daemon, the host on which the process runs experience a high load and, for that, the process fails to respond in time, etc. In order to cope with these problems we introduce a second layer of functionality in the architecture. In order to increase the level of confidence, at various moments of time the detectors exchange information between them, using a gossip-based approach, each one informing other detectors of their current knowledge of suspected processes. Upon receive of such an update a failure detector updates the local suspicious levels. For example, a process wrongly suspects a process of being failed because it does not have, from some moment of time forward, a direct network connection with that particular process. However, it further receives updates from another failure detection daemon in which he is informed that the process is still alive as it responds well to the second detector.

The gossip-based approach ensures that the system is able to detect errors such that: a failure detection service fails to directly communicate with a monitored process, there is an increased background networking traffic that results in an increased time needed for the process to respond back, etc. At various moments of time the detectors exchange information between them, using a gossip-based approach, each one informing other detectors of their current knowledge of suspected processes. Upon receive of such an update a failure detector updates the local suspicious levels.

Between clusters the information is propagated using specially designated failure detection services, located at the border of the cloud. In order to minimize the number of exchanged messages, the information is propagated only on requests coming from upper-level applications. This is motivated by the current functionality of large scale distributed applications. If we refer to Grid systems, a distributed application is decomposed into several tasks that are scheduled for execution using batching systems in one (or few) clusters. This is because tasks generally communicate mostly between them or with certain localized services and, by scheduling the tasks of an application inside one cluster (or a small number of closely situated clusters), the communication delays are minimized. For large scale distributed application the preferred way is to use a meta-scheduler, but even in this case an application is scheduled for execution in several clusters (if the application is big enough) that are closely localized. In case of cloud computing, an application is generally executed inside one cloud, which in our approach will constitute one cluster. Based on these observations, it is sufficient to let a failure detection services monitor only other services located inside the same cluster.

When an application or process needs information regarding the suspicion level that a certain process from another cluster failed it issues a request to a local failure detection service. This service then forwards the request to a border failure detection service from the second cluster (in our approach we use a DNS-based naming approach for finding border detectors) and obtains the value that he then sends back to the application. This approach has two major advantages. By using a propagation-on-request approach we minimize the number of messages being exchanged. Then, by monitoring only processes located in the same clusters, the probability of failure detections due to long delivery delays is reduced. Also, a solution in which each failure detector service maintains states regarding all the processes inside a distributed system is not practically feasible because the failure detection services run on hosts having limited amounts of resources (memory, cpu, etc.). The proposed failure detection architecture also scales well because each module is responsible with keeping only a small list of confidence levels for several processes.

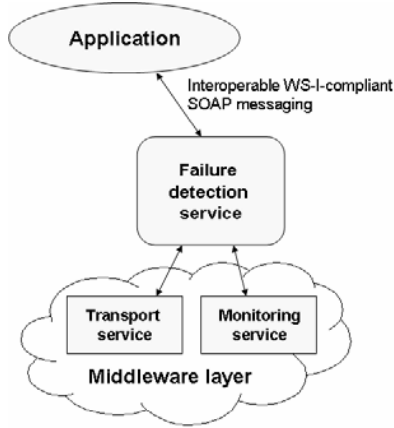

Fig. 3. The interoperability between the failure detection service, application and various middleware services.

Finally, the last layer is represented by the service capabilities being provided to various applications running on top of the large scale distributed system. As in case of accrual failure detectors (Defago, *et al*, 2005), we provide a complete decoupling between monitoring and interpretation. The failure detection architecture follows the SOA approach, applications being able to send requests regarding current suspicious levels of failures for certain processes from the failure detectors services using a standardized service approach. Also, this approach has the advantage of coping well with various existing service-based middleware platforms, providing several functionalities as presented in Figure 3.

The solution is designed to interact with various services provided by a possible underlying middleware services. For example, in order to query and obtain the confidence level of some process from another cluster a failure detector can use a transport service provided by the middleware (such as the GridFTP service provided by the Globus Toolkit), and in order to obtain accurate results on the reasons for a process failure it could interact with a monitoring service (such as MDS4 service provided by Globus Toolkit).

The architecture is designed to scale well and provide timely detection. For that, we combine the advantages of several proposed failure detection solutions. We believe that, in order to cope with the large scale nature of today's distributed systems, a failure detector must scale well and also the probability of false detections must not be influence by the number of monitored processes. For that, a gossip-based protocol provides several advantages (a formal demonstration is provided in (van Renesse, 1998): the probability that a member is falsely reported as having failed is independent of the number of processes; the algorithm scales in detection time and in network load, and for large networks the bandwidth used in the subnets is approximately constant.

We combine these properties with those introduced by the accrual detectors. Such detectors provide a lower-level abstraction that avoids the interpretation of monitoring information (see Figure 4).

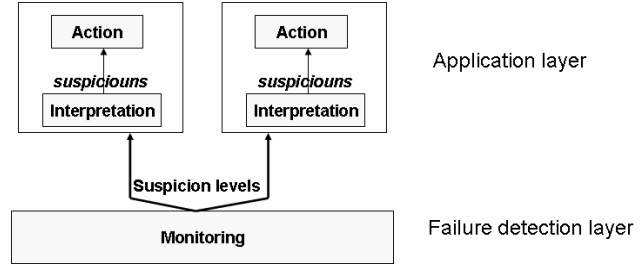

Fig. 4. Structure of the accrual failure detectors. Monitoring and interpretation are decoupled. Applications interpret a common value based on their own interpretation.

Some value is associated with each process that represents a suspicion level, which is then left to the application to be interpreted. In this way a real-time application could take quicker decision on processes being considered failed, while application requiring a high-level of confidence in their decisions (such as a data warehouse synchronization service) might require higher level of confidence that a process really failed. By setting an appropriate threshold, applications can then trigger suspicions and perform appropriate actions. Alternatively, applications can directly use the value output by the accrual failure detector as a parameter to their actions.

## 5. IMPLEMENTATION DETAILS

The building block for implementing the failure detection monitoring capability is the accrual detector construction (Defago, *et al*, 2005). An accrual failure detector outputs values from a range $R = (R^+)^\Pi$ (infinite range). In their case history is defined as $H(q,t)(p) = sl_{qp}(t)$. The failure detector module outputs non-negative real values, with each value corresponding to a process and representing the current suspicion level of that process.

As such, the monitored process $p$ sends heartbeats at regular intervals to the monitoring process $q$. Upon a query, the detector at $q$ simply returns the time that elapsed since the reception of the last heartbeat. Unlike previous solutions, we assume that processes can fail by crashing permanently, but also that they can only experience temporary crashing, due to high loads for example (in practice, a process that can not respond for a certain amount of time is also considered failed, since a process can not use any functionality provided by the process). Informally, if $p$ crashes, it stops sending heartbeats, and this triggers an increase on a suspicion level associated with that process. The suspicion level function satisfies the following two properties (Defago, *et al*, 2003):

Property 1. (Accruement) If process $p$ is faulty, then eventually, the suspicion level $sl_{qp}(t)$ is monotonously increasing at a positive rate.

Property 2. (Upper bound) If process $p$ is correct, then the suspicion level $sl_{qp}(t)$ is bounded.

The first property translates into the following. When $p$ stops sending heartbeats the suspicion level associated with it by process *q* increases forever. In contrast, if *p* is correct, it is possible to compute an upper bound on the maximal time elapsed between any two consecutive heartbeats (property 2), based on the characteristics of the execution.

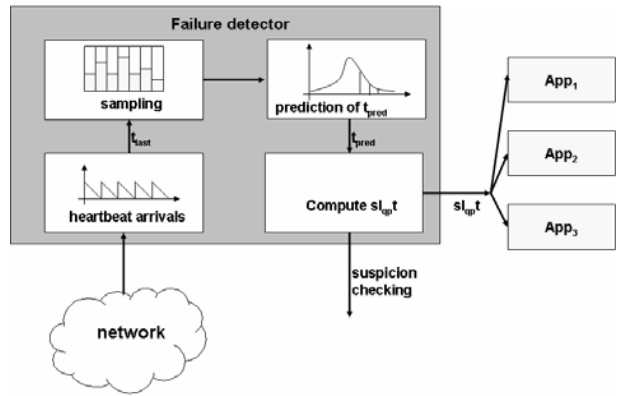

Fig. 5. Information flow in the proposed implementation of the failure detector.

Our implementation of the heartbeat accrual failure detector works as follows. Each heartbeat that was not received contributes partly to the suspicion level of the failure detector. The contribution of a heartbeat *H* increases from 0, meaning that *H* is not yet expected, to 1, meaning that *H* is considered lost. The suspicion level is calculated as a sum of all contributions. But, unlike previously other existing implementations of accrual detectors, the suspicious level in this case is not computed only from the local heartbeat contributions, but also from contributions received from other failure detectors located in the local cluster.

The contribution function is computed as follows. Each failure detector maintains a local suspicious level value $sl_{qp}(t)$. The heartbeat messages are sampled by the detector in order to estimate the time when the next heartbeat is expected to arrive. For that the detector can use any of several prediction methods. The predicted value for the next arrival of a heartbeat message is further used for computing $sl_{qp}(t)$:

$$sl_{qp}(t) = \max\{0, \log_{10}(t_{now} - t_{pred} + 1)\} \quad (1)$$

This means that, as time passes and the heartbeat message fail to arrive, $sl_{qp}(t)$ will come closer to 1. This strategy is described in Figure 5.

We next evaluated the accuracy of several prediction algorithms using a Java class for each method. The program runs as a background thread providing real time prediction for sampling of the heartbeat values.

*Simple Moving Average.* A simple prior moving average (SMA) is the unweighted mean of the previous *n* values. For example, a 5-minute simple moving average of the heartbeat is the mean of the previous 5 sampling values. Considering the values are $L_t, L_{t-1}, ..., L_{t-4}$, the formula is

$$SMA = \frac{L_t + L_{t-1} + L_{t-2} + L_{t-3} + L_{t-4}}{5} \quad (2)$$

While it is easy to implement and requires no additional overhead, SMA's performance is marginally satisfactory. By the nature of the calculation, this algorithm produces results that are both delayed and dampened. The algorithm has a tendency to flatten local peaks as a result of the averaging function, but the result generally follows the real trends.

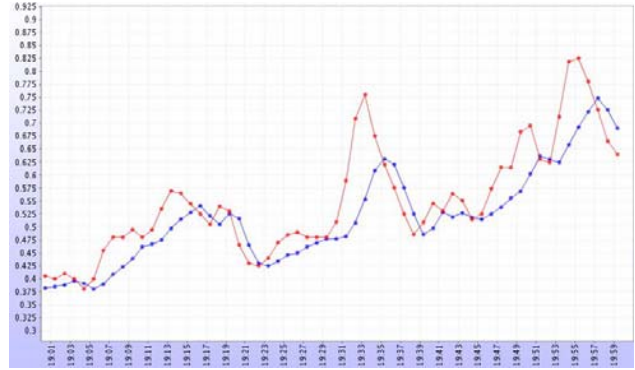

Fig. 6. Heartbeat interarrival real times (red) and predicted (blue) values using a restricted moving average.

*Restricted Moving Average.* Experimental results have shown that the behavior of the simple moving average prediction algorithm is not the desired one. The predicted value is sometimes an increasing value, and the real value is actually decreasing. To eliminate this behavior, a restriction on the moving average algorithm has been introduced. If the predicted value is higher than the last value, the predicted value will become the last real value that has been provided. When using this algorithm, the peak sensitivity problems persist and the restriction itself is a big source of errors, especially in the case of fast, high amplitude variations. In some cases the predicted values and the real ones show opposite trends (e.g. the real value increases but the predicted value is lower than the previous one).

*Weighted Moving Average.* A weighted moving average (WMA) is an average that has multiplying factors to give different weights to different data points. The weights are decreasing arithmetically as the values are older in time. In an n-value weighted moving average, the last value has weight n, the value before the last has weight *n - 1*, and so on.

$$WMA = \frac{5L_t + 4L_{t-1} + 3L_{t-2} + 2L_{t-3} + L_{t-4}}{5 + 4 + 3 + 2 + 1} \quad (3)$$

Although this prediction algorithm is giving extra weight to more recent data points, the predicted values are not very accurate. The reason is that the recent past may not offer sufficient information with regard to the next value of the monitored parameter. The weighting procedure assumes that more recent data is more significant, which in some cases may not be the case, for example for rapid fluctuations.

*Exponential Moving Average.* In an exponential (weighted) moving average (EMA) algorithm, the applied weights are decreasing exponentially. In this way, a more recent heartbeat arrival time is given much more importance than the weighted moving average algorithm. In the same time the algorithms does not discard older observations entirely. The constant smoothing factor is the degree of weighing decrease and is a number between 0 and 1. If the value at the time t is $L_t$ and the vales assigned to EMA at the same time is $S_t$, then $S_1$ will be undefined and $S_2$ will be initialized as the average of the first 5 values. The formula for calculating the exponential moving average at any time periods $t \geq 2$ is

$$S_t = \alpha \times L_{t-1} + (1-\alpha) \times S_{t-1} \qquad (4)$$

Depending on the constant $\alpha$, older values have more or less importance in the sum. If the smoothing factor is higher, the older observations are discounted faster. While the performance is generally better than the one expected from a simple moving average algorithm, this method fails to produce very accurate results when there is a significant difference between values at consecutive time points. Good results can be achieved by tuning the smoothing factor, if the general behavior of the signal is known. For completely random interarrival times, the results are only slightly better than the ones produced by the moving average technique, with the greatest error being produced mostly when the signal varies abruptly after a period of little or no change.

In terms of behavior, when the network is stable, i.e., few messages are lost, only one single heartbeat contributes to the suspicion level significantly, and thus the suspicion level reflects the contribution function. If the contribution function adapts well to the variability in arrival times (as in case of using an exponential moving average function), so will the applications using the failure detector. On the other hand, when the network is unstable with a lot of message losses, or if the monitored process crashes, contributions for all missed heartbeats but one will likely be close to 1. In this case, the failure detector will give a count of missed heartbeats, and the shape of the contribution function will be nearly irrelevant. In order to cope with that situation, we introduced another construction.

The construction is based on the use of a gossip-style approach. In this approach, a failure detector forwards information to other randomly chosen members of the group. Each member occasionally broadcasts its list in order to be located initially and also to recover from network partitions (van Renesse, *et al*, 1998). In the absence of a broadcast capability, the network could be equipped with a few gossip servers, that differ from other members only in that they are hosted at well known addresses, and placed so that they are unlikely to be unavailable due to network partitioning. This step ensures that the detectors know at least a subset of the entire group of failure detectors running in the local cluster (detectors themselves can fail, for this reason we state subset in this statement).

Whenever the suspicion level $sl_{qp}t$ reaches a certain threshold value *TV*, the failure detectors randomly selects several other failure detectors (for example, by tossing a weighted coin (van Renesse, *et al*, 1998) and sends its current value. It does not continue to increase the suspicion level computed using the accrual algorithm previously described until a certain condition occurs.

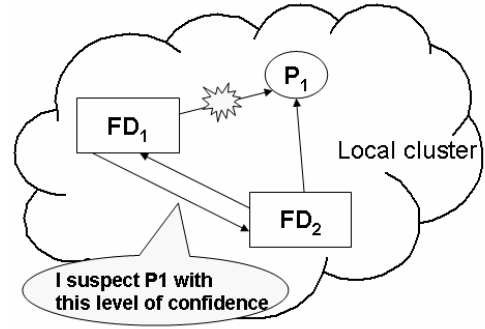

Fig. 7. The gossip-based information flow of local suspicious level values $sl_{qp}(t)$.

The gossiping has the role of eliminating false positives. A failure detector *q* sends out a message saying he suspects process *p* of being failed. Another failure detector *q'* eventually returns back an answer (or submits a message that its suspicion level for the same process crossed the threshold *TV*) containing its currently computed suspicion level $sl_{q'p}t$. At any moment of time the current suspicion level is considered to be the minimum value from the set of values obtained in this algorithm (the values received from other detectors, together with the locally computed suspicion level value). At some time a failure detector might receive again a heartbeat message from the suspected process. In this point the suspicion level starts to decrease, meaning the value again crosses the suspicion level *TV*. Again, it selects a random number of failure detectors and sends a message containing the current $sl_{qp}t$ value.

As demonstrated in (van Renesse, *et al*, 1998), this protocol does not impose a significant amount of load on the network bandwidth and is resilient against network partitions. However, unlike a basic gossip-based failure detector, it does not suffer from the disadvantage that failures have a negative influence on the number of rounds needed for information to be disseminated through the system, and hence on failure detection times across subnets and domains (because each detector already know a local level of confidence previously computed).

In order to better understand the behavior of the presented failure detection approach we analyze two cases. Assuming two detectors monitor the same process. When a link fails between the first detector and the process the value $sl_{qp}t$ will increase from 0 to 1. In the same time the suspicion level, in case of the second detector, will remain somewhere around 1. When the two detectors exchange information they will result in both knowing that one process still reaches the process and, thus, the process is still alive.

Next, we consider the case when a process runs on a workstation that becomes increasingly loaded. In this case the error is transient. All failure detectors will correctly suspect the process of being failed, as it will not send heartbeats anymore. When the workstation becomes free again the process will eventually start sending heartbeat messages. In this case the first detector that receives a heartbeat message will inform the other the process is alive, thus the time taken by the detectors to correctly detect the good health of the process is reduced compared to other existing detection approaches.

## 5. CONCLUSIONS

As society increasingly becomes dependent of distributed systems (Grid, P2P, network-based), it is becoming more and more imperative to engineer solutions to achieve reasonable levels of dependability for such systems. Failure detection constitutes a fundamental abstraction for fault tolerant distributed systems.

In this paper we presented a robust failure detection architecture that combines the power of existing approaches: fast propagation of information as offered by gossip-based failure detection approaches together with the decoupling of monitoring and interpretation as offered by the accrual failure detection solutions. The solution is based on several prediction functions and a new alternative of computing the contribution function. The approach has several advantages, among which we mention a better estimation of the interarrival times of heartbeat messages and an increase level of confidence in the suspicions of processes being lost.

The approach considers both the various networking conditions of large scale distributed systems and the different QoS detection requirements coming from various applications. In our approach the interpretation of the suspicion level is left to the distributed application using it. In this way multiple applications, having different QoS requirements, use the same failure detectors in different ways. The application could take either conservative (slow and accurate) or aggressive (fast, but inaccurate) decisions.

In order to cope with the large scale and dynamic nature of contemporary distributed systems, our solution considers a clustering of the detection functionality, with detectors being responsible with only subsets of the entire monitored processes of the underlying large scale distributed system.

In order to cope with transient errors, but also with the failure of network links we presented a solution in which we combined the detection function with an approach in which the detector exchange information using a gossip-style approach. The solution was further extended with a hierarchical information dissemination approach, in which information belonging to other clusters can be propagated from cluster-to-cluster when requested by applications. In this way, our solution has also the advantage of imposing a limited number of messages being exchanged in the network, mostly only on local levels, such that to be as non-intrusive as possible regarding the functionality of the entire distributed system.

In order to be of better use, the proposed failure detection offer capabilities to various applications as a service. This means that the framework can be easily incorporated in various existing distributed systems, and also can use the capabilities offered by various middleware architectures (such as it could use the transport capabilities offered by a transport service or can adapt the suspicion levels based on monitoring results obtained from a monitoring service, or can output current suspicion levels in a monitoring repository).

In the future, we aim to fully deploy this solution in various existing system and compare the obtained performances against various existing solutions. We also plan to extend the architecture in order to include not only detection capabilities, but also means to allow application to automatically asses various recovery and masking (redundancy) mechanisms.

## REFERENCES

Chandra, T.D., Toueg, S. (1996). Unreliable failure detectors for reliable distributed systems. *J. ACM*, *43(2)*, pp. 225--267. ACM.

Defago, X., Hayashibara, N., Katayama, T. (2003). *On the Design of a Failure Detection Service for Large-Scale Distributed Systems*. Intl. Symp. Towards Peta-bit Ultra Networks (PBit 2003), pp. 88--95, IEEE Computer Society.

Defago, X., Urban, P., Hayashibara, N., Katayama, T. (2005). *Definition and Specification of Accrual Failure Detectors*. 2005 International Conference on Dependable Systems and Networks (DSN'05). IEEE Computer Society.

Stelling, P., Foster, I., Kesselman, C., Lee, C., von Laszewski, G. (1998). *A fault detection service for wide area distributed computations*. 7th IEEE Symp. on High Performance Distributed Computing, pp. 268—278. IEEE Computer Society.

van Renesse, R., Minsky, Y., Hayden, M. (1998). *A gossip-style failure detection service*. Middleware'98, pp. 55--70, IEEE Computer Society.

Fetzer, C., Raynal, M., Tronel, F. (2001). *An adaptive failure detection protocol*. 8th IEEE Pacific Rim Symp. on Dependable Computing, pp. 146--153, IEEE Computer Society.

Chen, W., Toueg, S., Aguilera, M. K. (2002). *On the quality of service failure detectors*. IEEE Transactions on Computers, 51(2), pp. 13--32, IEEE Computer Society.

Bertier, M., Marin, O., Sens, P. (2002). *Implementation and performance evaluation of an adaptable failure detector*. Intl. Conf. on Dependable Systems and Networks (DSN'02), pp. 354--363, IEEE Computer Society.

Hayashibara, N., Defago, X., Katayama, T. (2004). *Flexible Failure Detection with K-FD. Research Report IS-RR-2004-006*.

Khanna, G., Laguna, I., Arshad, F.A., Bagchi, S. (2007). *Stateful Detection in High Throughput Distributed Systems*. 26th IEEE Intl. Symp. on Reliable Distributed Systems, pp. 275--287, IEEE Computer Society.